# Are Quantum States Subjective?

Rajat Kumar Pradhan*
Rajendra College, Bolangir, Odisha, India-767002.

**Abstract**

The subjective nature of the quantum states is brought out and it is argued that the objective state assignment is subsequent to the subjective state of the observer regarding his state of knowledge about the system. The collapse postulate is examined in detail to bring out the inherent subjectivity of the quantum state. The role of doubt and faith in quantum state assignment is examined.

Key words : Quantum measurement, state vector reduction, collapse postulate, superposition

PACS numbers: 03.65 Ta, 03.65 Ca

## 1. Introduction

The problems of the interpretation of quantum theory [1-13] have long since engaged the attention of physicists and philosophers alike, right from the days of its inception. The main problem discussed by Schrödinger[3] through the celebrated 'cat paradox' is the 'probability interpretation' which makes the theory indeterministic. The other problem is our inability to obtain simultaneous definite values for conjugate observables which is brought to focus by the 'EPR-paradox'[2]. These two paradoxes have their myriad variants [1, 4, 9] and many have been the proposals put forth in the literature to solve them including the introduction of the conscious observer [8] into the very heart of the discussions.

Interestingly enough, these basic problems notwithstanding, the theory has had monumental success not only in describing physical phenomena but also in the description of phenomena in some other areas [11, 12, 13] as well like sociology, psychology, philosophy etc. The superposition postulate following originally from the linearity of the Hamiltonian has proved a handy tool in describing the dynamics of systems having many possible states. But, the link of quantum theory with the conscious observer still remains shrouded in mystery [14, 15, 16] and the nature of the quantum state and of the measurement process remain open to ever-newer interpretations [17].

Traditionally, QM is introduced through the following postulates:

- **Postulate-1:** (*Hilbert space*): The normalized quantum state $|\Psi\rangle$ which contains all the information about the system is an element of a linear vector space, the Hilbert space, spanned by a set of appropriate basis vectors for the system.
- **Postulate-2:** (*Hermitian Operators*): An observable is represented by a hermitian operator **A** having real eigenvalues. The basis vectors are then the simultaneous eigenstates of the complete set of mutually commuting observables.
- **Postulate-3:** (*Expectation value*): The measurement results are interpreted statistically for a large number of identically prepared systems through the postulate of the 'expectation value':
$$\langle \mathbf{A} \rangle = \langle \Psi | \mathbf{A} | \Psi \rangle \qquad \ldots \ldots \ldots (1)$$
- **Postulate-4:** (*Time Evolution*): The time evolution is governed by the Schrodinger equation:
$$\mathbf{H} |\Psi\rangle = i\hbar (\partial |\Psi\rangle / \partial t) \qquad \ldots \ldots \ldots (2)$$

and, finally,

- **Postulate-5:** (*The collapse postulate*): Upon measurement of the observable **A** with eigenvalue $a_i$, the original (linear superposition) state $|\Psi\rangle = \Sigma_i c_i |a_i\rangle$ collapses to the state $|a_i\rangle$ with probability $P_i$ given by :
$$P_i = |c_i|^2 = c_i^* c_i = \langle \Psi | a_i \rangle \langle a_i | \Psi \rangle \qquad \ldots \ldots \ldots (3)$$

Where, $c_i = \langle a_i | \Psi \rangle$ is the probability amplitude for the system to be found in the state $|a_i\rangle$.

This last has been the center of attention since von Neumann's original work [8] bringing the conscious observer into the foundations of quantum theory.

Many have been the proposals to keep out the observer and to safeguard the objectivity of physics [18], but the most recent works [11,12,14,17] have consistently pointed in the direction of an inevitable entry and a decisive role of the observer in the quantum measurement process.

In the present work, it is argued that the state of the system is primarily a subjective assignment reflecting the observer's state of knowledge about it. The distinction between micro- and macro- systems made in the decoherence program [19, 20] is rather artificial, since in no case our knowledge of the state is ever definite. This is primarily because, in reality, as explained by Dirac [21] we never measure any exact eigenvalue, but rather a value having a spread (error or uncertainty) depending on (a) the system, (b) its interaction with the environment, (c) the observable measured, (d) the measuring apparatus and, possibly, as required in QM, (e) the observer.

Thus the assignment of an exact eigenvalue neglecting the spread (and an exact eigenstate) is therefore a subjective affair even for microscopic systems like the H-atom and is only theoretically justifiable as a convenient idealization. If this were not the case, and if indeed, we had really exact eigenvalues and states we would never have heard of fine structures in the spectral lines of atoms. It is clear that there is always an inevitable superposition of states within the spread irrespective of whether the system is microscopic or otherwise. The only difference is that for microscopic systems we are forced to accept the superposition of states while for the macroscopic ones we habitually disregard them as of no consequence.

In section-2, the subjective nature of the quantum state is brought out by a detailed analysis of the collapse process in a quantum measurement. In section-3, the central role played by the observer's doubt and faith in state assignment is considered before concluding in section-4 with a discussion of the main results.

## 2. The Subjective nature of quantum states

Consider a system described by the Hamiltonian **H**. In the measurement of an observable **A**, the general pre-measurement state $|\Psi\rangle = \Sigma_i c_i |a_i\rangle$ collapses to the eigenstate $|a_i\rangle$ with probability $|c_i|^2$. If $[\mathbf{H}, \mathbf{A}] = 0$, then the final state will be the simultaneous eigenstate of **H** and **A**. If they are incompatible, then the eigenstate $|a_i\rangle$ will no longer be an energy eigenstate but will be a superposition of the energy eigenstates $|e_n\rangle$ defined by :

$$\mathbf{H} |e_n\rangle = \varepsilon_n |e_n\rangle \qquad \ldots \ldots \ldots (4)$$

Thus, the assignment of probabilities to the occurrence of eigenvalues is a consequence of our ignorance of the state before measurement, since just after the measurement the state is definitely known to be $|a_i\rangle$. A definite knowledge gets rid of the probabilities. Prior to the measurement, the observer must know the Hamiltonian and, if possible, the complete set of mutually commuting observables. This, itself requires some knowledge of the system's constitution, which can only be obtained through experiments.

Both classical as well as quantum probabilities are due to a lack of complete knowledge of the state or of the measurement outcome, but the difference lies in the fact that the classical probability is the squared modulus of the corresponding quantum probability amplitude. This leads to the appearance of quantum interference terms in the calculation of probabilities for a superposed state.

Probability is a measure of the inadequacy in our knowledge and that the quantum state is our subjective assignment depending on our state of knowledge, the notion of objective state vector reduction (independent of the observer) due to coupling with the classical macroscopic environment in the decoherence program then becomes an assumption which is based on the presumed exactitude of classical knowledge of the macro-state of the environment. As long as the classical system has an (inevitable) spread in the values of observables, so long the microstate (of the quantum system coupled to the classical one) corresponding to the same observable will be a superposition. The probability amplitudes are either calculated theoretically, in which case we must have performed measurements yielding sufficient knowledge of the constitution of the system and its coupling to the environment, or, are measured experimentally as square root of the corresponding probabilities.

We wish to prove that ***the reduction or definite knowledge is registered, exactly where there was the superposition or the inexact knowledge, i.e. in the particular observer's consciousness.***

Suppose, two observers $o_1$ and $o_2$ share the same knowledge of the pre-measurement state $|\Psi\rangle = \Sigma_i c_i |a_i\rangle$. Now, let's say, observer $o_1$ makes a state-reducing measurement, collapsing it objectively to a particular eigenstate $|a_1\rangle$ and then subjectively coming to the confirmation of the post-measurement state being truly $|a_1\rangle$ upon the observation of the pointer state of the measuring device.

Now, as long as $o_2$ does not know about the reduction, he will continue to assign the same old (pre-measurement) state $|\Psi\rangle$ to the system in spite of the objective reduction as well as the subjective knowledge in $o_1$'s consciousness about its reduction to $|a_1\rangle$. Since the first quantum postulate states that the wave function contains complete information about the system, $o_2$ can be said to have incomplete or, at least, partially complete information compared to $o_1$.

But, if both of them make a large number of fully independent measurements on such identically prepared systems, they will doubtless verify their common knowledge $|\Psi\rangle$ of the premeasurement superposition state $|\Psi\rangle$.

However regarding the outcome of a single measurement by each observer, there are three distinct possibilities that arise for $o_2$ to assign a state to the same system following $o_1$'s measurement:

**I.** *$o_2$ knows neither about the measurement nor about the outcome*: Without any knowledge of $o_1$'s measurement, if $o_2$ makes a measurement on the same system shortly after $o_1$, then she will definitely get the state $|a_1\rangle$. This can be predicted with certainty by $o_1$, but $o_2$, without knowing about $o_1$'s interaction with the system, will interpret it as having come about due to the collapse, which had a prior probability of $|c_1|^2$.

**II. *$o_2$ knows about the measurement, but not about the outcome*:** She will assign a new state $|\Psi'\rangle$ to the composite system (original system plus apparata) because of her enhanced knowledge of the interaction of $o_1$'s measuring apparatus with the system:

$$|\Psi'\rangle = \Sigma_i c_i |a_i\rangle |d_i\rangle \quad \ldots \ldots \ldots \quad (5)$$

Where $|d_i\rangle$ are the states of the measuring device commonly called the 'pointer sates'. Now, It is this new state $|\Psi'\rangle$ that will collapse in $o_2$'s consciousness to $|a_1\rangle$ with probability $|c_1'|^2$, if she makes another measurement quickly to ascertain the state.

**III. *$o_2$ knows about the measurement as well as the outcome*:** In this case naively, and habitually, we take it for granted that $o_2$ will assign the state $|a_1\rangle$ to the system since it is the common objective knowledge of the outcome of a scientific measurement.

However, here also caveats are there since, only if $o_2$ has *faith* in the outcome, then and only then the state collapses in her consciousness with the rise of definite knowledge, and no further measurement is required for her to assign the state. But if *doubts* arise regarding the outcome, then the lack of definite knowledge in $o_2$'s consciousness will require a further measurement to ascertain the state of the system which may or may not tally with $o_1$'s result obtained previously, depending on whether the state is a stationary one or not, and also, on the time elapsed in case it is allowed to evolve into a superposition again.

Thus, in all the three cases we see that in general the two observers do not agree with each other in regard to the state of the system even after a measurement by one of them. This is called 'weak objectivity' in a recent essay by d'Espagnat [22].

We therefore turn now to the role of doubt and faith in quantum physics and its applications in particular so that a better appreciation of the role of the conscious observer can be obtained.

## 3. Role of doubt and faith in quantum physics

It is clear from the above discussion that, a *doubt* can be represented as a superposition and *faith* as a reduction or collapse. If, for any reason whatsoever, there is doubt in $o_2$ regarding the results of $o_1$'s measurements, then her lack of definite knowledge is as good as a superposition, and she needs to perform a second measurement immediately to ascertain the state for herself and thereby have the superposition collapse. If on the other hand, she puts her faith in the results of the first measurement, then the state automatically collapses in her consciousness leading to definite knowledge. It is to be noted that sometimes, for some reason or the other, we may have doubts regarding the validity of our own observations and in that case the same reasoning as above also applies i.e. a superposition occurs and a fresh measurement is called for to effect a collapse.

However, *doubt* and *faith* are aspects or functions of consciousness of the observer and not of the system under consideration! Also, this shows that our collective unanimity of *faith* in the truth of the outcome of a certain experiment or of an analysis is what grants *objectivity* to phenomena.

Interestingly enough, Physics has benefitted from both, *doubts* and *faiths*, in a big way! For example, Einstein *doubted* the absoluteness of space and time held by Newton and propounded the theories of relativity, while Feynman had *faith* in Dirac's 'mysterious' remark on the weights for the paths and was led to the path integrals. On the experimental side, Planck had *faith* in the experimental curve obtained for radiation intensity from a black body and this led him to the quantum hypothesis to get a neat fit to it. Had he *doubted* the curve, he would never have proposed the quantum hypothesis!

Einstein's *doubts* about the validity or veracity of the 'uncertainty principle' led him to propose many *gedanken* experiments during the Solvay conference of 1927, which all were refuted by Bohr with his *faith* in the validity of the same. The most famous and fruitful one in the series of these proposed experiments, was the celebrated EPR- paradox. Also Schrodinger's 'cat paradox', the de Broglie-Bohm pilot wave theory and the hidden variables theories (HVTs) similarly arose out of such *doubts* about the validity of the probability interpretation.

An experiment is performed either to gather new data, or to verify or refute a claimed experimental fact or theoretical prediction, or to improvise upon an earlier result in regard to its accuracy. Both *doubt* and *faith* may be seen to be lurking behind all these sincere efforts, though it has to be admitted that these *doubts* and *faiths* in scientific pursuits are more of the reasoned and rational category. In many instances, these are seen to be of the intuitive category also (like Einstein's doubt about the absoluteness of space and time which led to Relativity).

The differential operators of Schrödinger's wave mechanics, the hermitian operators of Heisenberg's matrix mechanics and the linear vector (Hilbert) spaces of Dirac's formulation are all acceptable to us, primarily because of our *doubt* (lack of definite knowledge) about the exact state of a system which provides the basis for the expansion postulate of quantum mechanics. Because, the premeasurement state of a quantum system cannot be known exactly, we accept these formulations as valid descriptions and then the probability interpretation naturally follows. Similarly also, the inability to ascertain the exact path that a system follows leads to the acceptance of Feynman's path integral formulation. Therefore, what subjectively is a *doubt* is construed (or reflected) objectively as a probability in a superposition of quantum states.

And, interestingly, what subjectively is a firm *faith*, mostly resulting from mathematical analyses, is often seen to be realized as an objective fact like in the case of the verification of so many theoretical predictions viz. the existence of the positron, the neutrinos, the $\Omega$- hyperon and also of the massive vector bosons $W^{\pm}$ and the $Z^0$ !

Now, knowledge or the lack of it ( i.e. *faith* or *doubt*) is a property of the consciousness and this is seen to be at the back of all our formulations of quantum mechanics. We can say that the quantum state usually ascribed to a system is a representation of our knowledge (or the lack of it) of the system. The quantum state is primarily in the observer's consciousness, and it represents the state of the particular observer's consciousness in respect of that particular system. The seeming objectivity comes about on account of the collective *faith* reposed in the ascribed state and it may be abandoned at any moment as and when 'subjective knowledge' contrary to the assigned state start coming to the surface leading to *doubts*, and then we would require fresh experiments to ascertain the state all over again.

## 4. Discussion and conclusion

The quantum state of a system has both objective and subjective representations. From the foregoing analysis it seems that we have to develop a more complete view of the measurement process as '*an arrangement of apparata that connect the subject and the object in a manner as desired or designed by the subject and the outcome has always a two way effect- it not only collapses the state of the object but also that of the subject*!' Now, which one collapses first? Objectively speaking, the information from the apparatus goes to the subject whereby she comes to know about the collapse and thus the subjective collapse is latter than the objective collapse.

However, from the subjective view, the premeasurement state was assigned by the subject basing on her knowledge of the possible states of the system. The state assignment thus proceeded from the subject to the object. After the measurement, the state assignment again follows from the subject to the object but not before the information from the apparatus reaches her, thereby collapsing her subjective state first. Then only, the definite knowledge of the collapse having taken place arises and she can say that the objective state has collapsed. Thus, the subjective collapse precedes the objective collapse as far as cognition of the collapse is concerned. As long as the subjective knowledge of the observer about the state remains uncertain, so long the state assigned by that particular observer will be a superposition, no matter what the state is for other observers. This corroborates the analysis made by Mould[23].

The above very simple analysis suggests that the quantum state is not an observer-independent assignment. In Rovelli's Relational Interpretation[24], the relative nature of the quantum states is taken in but the formulation falls short of granting conscious status to the observer. For Rovelli, any interaction of anything with the system qualifies to be called a measurement. We are of the view that "*the interaction of the apparata with the system makes up only the objective part of the measurement and the observation of the result by the conscious observer is the indispensable subjective part needed to complete the process of measurement, leading to the rise of definite knowledge.*"

In this work, we have argued that both the subject and the object are affected by a measurement and that there is a collapse in both. It seems that we have to admit the fact that the quantum state not only has a subjective aspect but also that the subjective and the objective states are like mirror images[25] of each other, when they come into being via measurement.

## Acknowledgements

The author wishes to gratefully acknowledge discussions with L. P. Singh and N. Barik on the role of consciousness and subjectivity of states in quantum measurements.